\documentclass[useAMS,usenatbib]{mn2e}
\usepackage{graphicx}
\usepackage{color}
\usepackage{soul} %%%\st{Hellow world}
\usepackage{url}
\usepackage{txfonts}
\usepackage{longtable}

\definecolor{myred}{rgb}{0.7,0.0,0.2}

\definecolor{myblue}{rgb}{0.0,0.2,0.7}

\definecolor{mygreen}{rgb}{0.2,0.7,0.0}

\title[WN/C star in M31]{The first transition Wolf-Rayet WN/C star in M31}
\author[M. Shara et al.]{Michael M. Shara$^{1}$\thanks{E-mail: mshara@amnh.org}, Joanna Miko{\l}ajewska$^{2}$, Nelson Caldwell$^{3}$, 
\newauthor Krystian I{\l}kiewicz$^{2}$, Katarzyna Drozd$^{2}$ and David Zurek$^{1}$ 
\\
$^{1}$Department of Astrophysics, American Museum of Natural History, Central Park West at 79th Street, New York, NY 10024, USA\\
$^{2}$N. Copernicus Astronomical Center, Bartycka 18, PL 00--716 Warsaw, Poland\\
$^{3}$Harvard-Smithsonian Center for Astrophysics, Cambridge, MA 02138, USA\\
}

\begin{document}

\date{Accepted  Received }

\pagerange{\pageref{firstpage}--\pageref{lastpage}} \pubyear{}

\maketitle

\label{firstpage}

\begin{abstract}
Three decades of searches have revealed 154 Wolf-Rayet (WR) stars in M31, with 62 of WC type, 92 of WN type and zero of transition type WN/C or WC/N.
In apparent contrast, about two percent of the WR stars in the Galaxy, the LMC and M33 simultaneously display strong lines of carbon and nitrogen, i.e. they are transition type WN/C or WC/N stars. We report here the serendipitous discovery of M31 WR 84-1, the first transition star in M31, located at RA = 00:43:43.61  DEC = +41:45:27.95 (J2000). We present its spectrum, classify it as WN5/WC6, and compare it with other known transition stars. The star is unresolved in Hubble Space Telescope narrowband and broadband images, while its spectrum displays strong, narrow emission lines of hydrogen, \mbox{[N\,{\sc ii}]}, \mbox{[S\,{\sc ii}]} and \mbox{[O\,{\sc iii}]}; this indicates a nebula surrounding the star. The radial velocity of the nebular lines is consistent with that of gas at the same position in the disc of M31. The metallicity at the 11.8 kpc galactocentric distance of M31 84-1 is approximately solar, consistent with other known transition stars. We suggest that modest numbers of reddened WR stars remain to be found in M31.
\end{abstract}


%==================================================================

\section{Introduction}

\subsection{Motivation}

Classical Wolf-Rayet (WR) stars display powerful, radiation-driven winds which are ejecting copious quantities of mass. These stars' spectra are dominated by strong emission lines of ionized carbon (the WC subtype) or nitrogen (the WN subtype), while helium lines are ubiquitous in all WR stars. The winds' peeling away of the outer layers of WR stars initially reveals the products of hydrogen burning (especially ionized nitrogen) in the WN stars, followed by the products of helium burning (carbon and oxygen) in the WC stars. This simple but elegant picture of massive star evolution, reviewed in detail in \citet{cro07}, predicts that there should exist a short-lived transition state between the WN and WC stages, when the emission lines of carbon and of nitrogen are visible simultaneously. These stars are thus valuable as laboratories where the last traces of the hydrogen burning products are being stripped to reveal the outer layers of helium-burning. Transition stars are designated as WC/N or WN/C, depending on the overall appearance of the emission-line spectrum; the element with stronger lines appears first \citep{con89}.

The models of \citet{lan91}, \citet{lan94} and \citet{mey94} explored the effects of semi convection, varying spin rates, metallicity and mass loss rates on the relative lifetimes and expected numbers of WR stars of different subtypes. The thickness of the transition zone, and hence the time during which a star displays both WN and WC characteristics, is predicted to increase with increasingly effective semi convection and more rapid rotation. The WN/C lifetime is predicted to decrease with increasing metallicity (and hence mass loss rate), as these strip away the transition zone faster. The observed number of transition WN/C stars relative to all WR stars is thus an important constraint and test of the predictions of the models of the late evolutionary stages of massive stars' evolution.

It is observed that the rarest of all Wolf-Rayet stars are the WO subtypes \citep{tra15}, with nine currently known, closely followed by the transition types. Only 12 of the latter are known in the Milky Way, the LMC, SMC, IC10 and M33 \citep{mor87,con89,sch90, bre99,cro03,mas14} out of a total of about 600 WR stars with spectra of quality high enough to distinguish their transition nature. This rarity, corresponding to about 2\% of all WR stars, empirically demonstrates that the transition time from WN to WC must be short - of order 10,000 years \citep{cro95}. 

%6 WN/C in MW when the total Galactic census was 158 stars.
%In the LMC, with 152 WR stars  2 are of type WN/WC.
%In M33 \citep{sch90,neu11} three of 206 WR stars are WN/WC objects, while 1 of the 26 WR stars in IC10 is a transition object 

The first narrowband imaging surveys for M31 Wolf-Rayet stars began three decades ago, looking for stars displaying strong emission lines of ionized helium, nitrogen and carbon \citep{mof83,mas86,mof87}. Currently the most sensitive and complete survey's \citep{neu12} estimate of the population of M31 Wolf-Rayet stars is 154, with 62 spectrographically-confirmed WC stars and 92 of type WN. This is far fewer than the $\sim 640$ WR stars currently known in our own Galaxy, which in turn is probably only $10\%$ of the Milky Way's total population \citep{sha09} (but see a lower estimate from \citet{ros15}). Not a single WN/C transition star has been reported in M31, while one might have expected 2\% $\times$ 154 stars $\sim 3$ stars to have been found. This total lack of M31 WN/C stars might be due to the high metallicity of M31's stars, or simply due to small number statistics. We report, here, the first transition WR star to be detected in M31, demonstrating that at least one such object does, in fact, exist there.

In Section 2 we describe the data and their reductions. The coordinates, images, observed and dereddened spectra, and classification of the new M31 WR transition star are presented in Section 3. We contrast and compare it with other WR transition stars in Section 4, and briefly summarize our results in Section 5.

\section{Observations and data reduction}\label{obs}

The WR star discussed in this paper was found as a ``by-product" of a spectrographic survey of M31 aimed at detecting and characterizing symbiotic stars (SySt) in that galaxy. 
Candidate SySt (and the WN/C star that is the focus of this paper) were chosen because they displayed strong H$\alpha$ emission in images of the publicly-available LGGS survey \citep{mas06}, and because they are quite red ($V - I \ga 2.0$). While all SySt display a strong H$\alpha$ line, they have little or no emission in the familiar forbidden lines of \mbox{[N\,{\sc ii}]}, \mbox{[S\,{\sc ii}]}, and \mbox{[O\,{\sc ii}]} \citep{mik14}. Instead, they show lines of high ionization potential such as \mbox{He\,{\sc ii}}.

The spectra themselves were obtained with the Hectospec multi-fiber positioner and spectrograph on the 6.5m MMT telescope \citep{fab05}.
The Hectospec 270 gpm grating was used and provided spectral coverage from roughly $3700-9200${\AA} at a resolution of $\sim5${\AA}. The observations were made
on the night of 17 November 2014, and were reduced in the uniform manner outlined in \citet{cal09}. The frames were first de-biased and flat fielded.  
Individual spectra were then  extracted and wavelength calibrated. Sky subtraction is achieved with Hectospec by
averaging spectra from "blank sky" fibers from the same exposures or by offsetting the telescope by a few arcseconds. 
Standard star spectra obtained intermittently were used for flux calibration and instrumental response. These
relative flux corrections were carefully applied to ensure that the relative line flux ratios would be accurate. The total exposure time was 5400 s.

Archival images of the field of the new WR star were downloaded from MAST, the Mikulski Archive for Space Telescopes, at the Space Telescope Science Institute. 
The field was observed for program 9794 (PI:Massey) on 2 December 2003 through multiple filters with the Hubble Space Telescope's Wide Field Channel of the Advanced Camera for Surveys. We also downloaded images of the LGGS from the Lowell Observatory's website, and carried out aperture photometry of every star in every image.

%\Table - emission lines 
\begin{table}
 \centering
  \caption{List of emission lines observed in the spectrum of the new WR transition star in M31, as well as their radial velocities ($\varv_{\rm rad}$), full widths at half maximum (FWHM), fluxes ($F(\lambda)$) and equivalent widths (EW).}\label{eml}
  \begin{tabular}{@{}llrrrr@{}}
  \hline
ID& $\lambda_{\rm obs}$\,[\AA] & $\varv_{\rm rad}^{\rm a}$  & FWHM$^{\rm a}$ & $F(\lambda)^{\rm b}$ & EW\,[\AA] \cr
\hline
\mbox{[O\,{\sc ii}]}\,3728$^{\rm c}$&3727.2&&1080&5.8&70 \cr
\mbox{O\,{\sc iv}}\,3811-34&3813.8&&1100&1.7&21 \cr
%[NeIII]3868.7&3862.5&-480&640&1.2&16 \cr
H$\gamma$\,4340.5&4337.4&-214& 590 &1.1&7 \cr
\mbox{C\,{\sc iv}}\,4659-60&4658.5&& 620&1.1&6 \cr
\mbox{He\,{\sc ii}}\,4685.7&4682.8& -186 & 2200&12.9&67 \cr
H$\beta$\,4861.3&4859.1&-136& 650 &2.8&13 \cr
\mbox{[O\,{\sc iii}]}\,4958.9&4957.5&-85& 700 &1.1&5 \cr
\mbox{[O\,{\sc iii}]}\,5006.9&5004.5&-144& 520 &2.2&10 \cr
\mbox{He\,{\sc ii}}\,5411.5&5408.0& -194 & 2400&6.3&25 \cr
\mbox{O\,{\sc v}}\,5494&5494.0&& 2100 &1.4&4.7 \cr
\mbox{C\,{\sc iii}}\,5696&5694.4&&1000&1.6&5.6 \cr
\mbox{C\,{\sc iv}}$^{\rm c}$\,5808 &5806.0& &2200&7.3&24 \cr
\mbox{He\,{\sc ii}}\,5875.6&5869.7&-301& 460 &0.8&2.5 \cr
\mbox{[N\,{\sc ii}]}\,6548.1&6543.7&-201 & 600 &5.4&18 \cr
H$\alpha$\,6562.8&6559.2&-165& 460 &20.4&67 \cr
\mbox{[N\,{\sc ii}]}\,6583.3 &6579.2&-187&500 &9.1&30 \cr
\mbox{He\,{\sc ii}}\,6678.2&6672.1&-274&1350 &2.3&8 \cr
\mbox{[S\,{\sc ii}]}\,6716.4&6712.8&-161& 450 &3.1&10 \cr
\mbox{[S\,{\sc ii}]}\,6730.8&6727.3&-156&450&2.3&7.6 \cr
\mbox{N\,{\sc iv}}\,7109&7107.2& &1700&9.0&24 \cr
\mbox{[Ar\,{\sc iii}]}\,7135.8&7130.0&-244& 380&1.8&5 \cr
\mbox{C\,{\sc iii}}\,7590$^{\rm c}$ & 7591.6 & & 1400 & 3.2 & 9 \cr
\hline
\end{tabular}
\begin{list}{}{}
\item $^{\rm a}$ In units of $\rm km\,s^{-1}$
\item $^{\rm b}$ In units of $10^{-17}\, \rm erg\, s^{-1}\, cm^{-2}$.
\item $^{\rm c}$ Blend of two components.
\end{list}
\end{table}

\section {The WN/C star}

\begin{figure*}
\resizebox{\hsize}{!}{\includegraphics{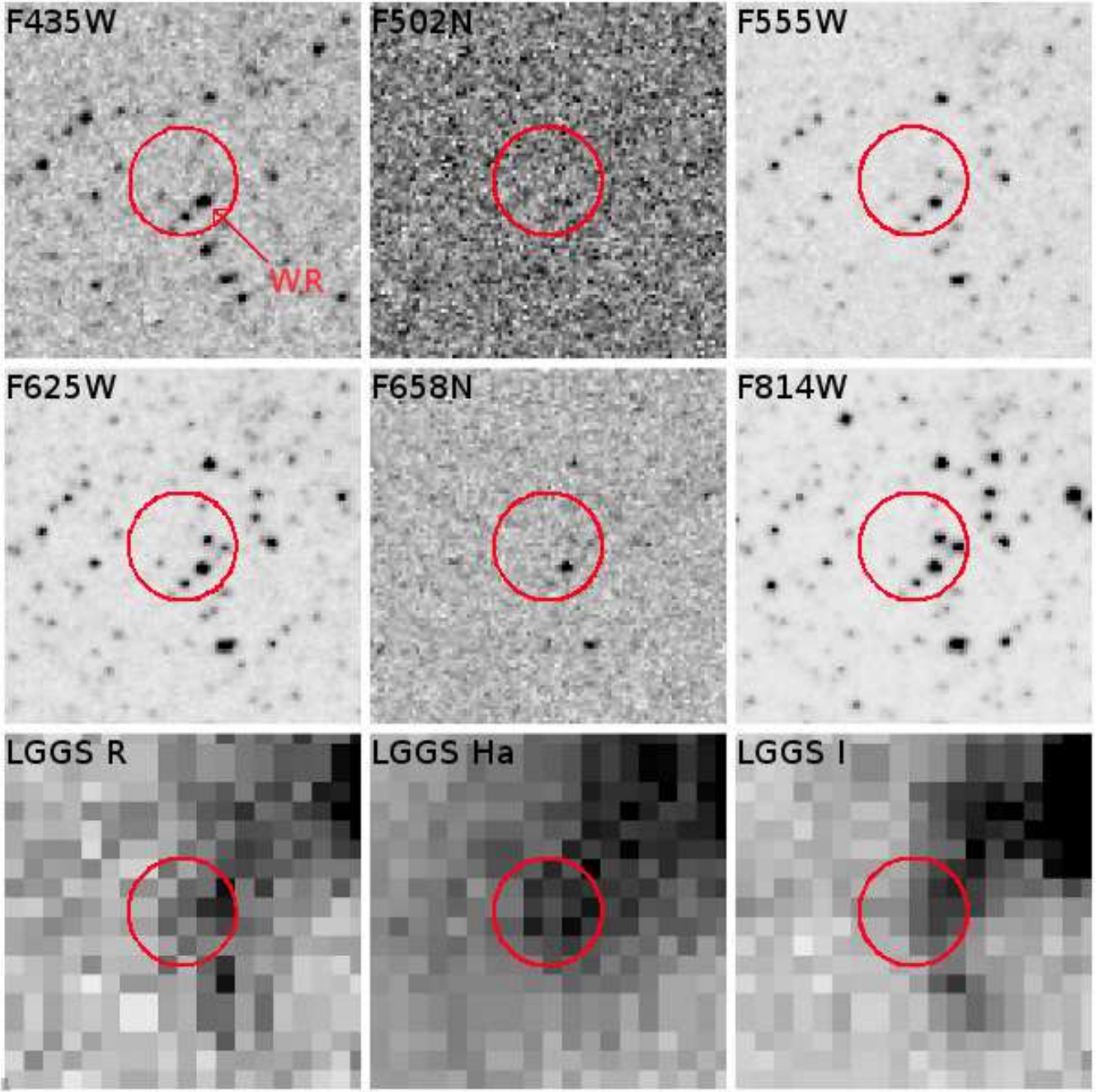}}
%\centerline{\includegraphics[width=0.99\columnwidth]{384433_R.eps}}
\caption{HST and LGGS images of the region near the new WR transition star (marked with the arrow) that we have detected in M31. N is up and E is left, and the scale of each image is 5$\times$ 5 arcsec. The red circles have 1.5 arc sec diameters, equal to the sizes of the fibers we used with the Hectospec spectrograph.}\label{FC}
\end{figure*}

The coordinates of the new WR transition star are 00:43:43.61  +41:45:27.95 (J2000). Following the naming convention for M31 WR stars introduced in \citet{san14}, we name it WR 84-1. It displays the following magnitudes and colors: $I =20.63$, $V-I = 2.08$, $R-I=1.00$, $m({\rm H\alpha}) - R = -1.01$,  resulting from our aperture photometry in the LGGS images. We present its finder chart in Fig.~\ref{FC}.

The new WR star (marked with the arrow in the F435W image) is the brightest object in all HST filters. It is the only object visible in the narrow F502N  and F658N filters, which basically represent the \mbox{[O\,{\sc iii}]} and H$\alpha$+\mbox{[N\,{\sc ii}]} images, respectively.

\begin{figure}
\centerline{\includegraphics[width=0.99\columnwidth]{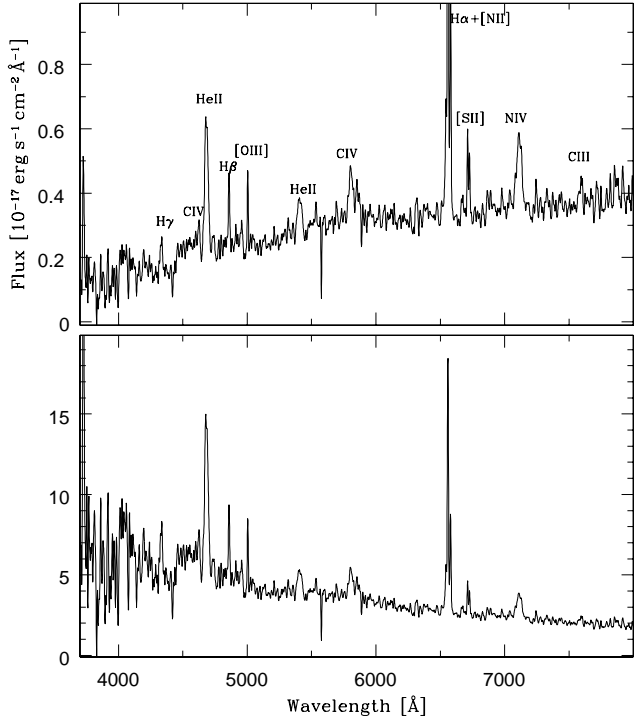}}
\caption{(Top) The spectrum of the new WN5/WC6 transition star we have detected in M31. The simultaneous presence of \mbox{C\,{\sc iv}}\,$\lambda5808$ and the even stronger \mbox{N\,{\sc iv}}\,$\lambda 7109$ emission lines demonstrate that the star is of type WN/C.
(Bottom) Same spectrum dereddened with E(B-V) = 0.9 (see text).}\label{spectra}
\end{figure}

The observed spectrum of M31 84-1 (normalized to $V=22.7$) is shown in the top portion of Fig.~\ref{spectra}. Strong and very broad emission lines of \mbox{He\,{\sc ii}}, \mbox{C\,{\sc iv}} and \mbox{C\,{\sc iii}} immediately identify the object as a WC star. Remarkably, even stronger than the \mbox{C\,{\sc iv}} lines is the very broad line of \mbox{N\,{\sc iv}} at 7103 {\AA}. In addition, narrow \mbox{H\,{\sc i}} Balmer, and forbidden  \mbox{[N\,{\sc ii}]}, \mbox{[S\,{\sc ii}]}, \mbox{[O\,{\sc ii}]} and \mbox{[O\,{\sc iii}]} emission lines of circumstellar origin are present  (see Sec. 3.1). The relative strength of the H$\alpha$ to the H$\beta$ emission lines suggests a reddening of approximately E(B-V) =0.9. We have accordingly dereddened the spectrum, and display it in the bottom portion of Fig.~\ref{spectra}. We list the star's emission lines' IDs, observed wavelengths from Gaussian line fitting, FWHM, line fluxes and line equivalent widths in Table 1. 

The simultaneous presence, as noted above, of strong lines of both \mbox{N\,{\sc iv}} and \mbox{C\,{\sc iv}}, as well as a range of ionization stages of C and He strongly suggest that M31 84-1 is a transition WR star. The \mbox{C\,{\sc iv}} 5808 {\AA} line strength of M31 84-1 is similar to those of the transition stars WR8, WR 98 and WR 153, while its \mbox{He\,{\sc ii}} emission lines are stronger. This is quantified in figure 5 of \citet{con89}, where transition stars are differentiated from WN stars with CIV emission lines in a plot of the log of equivalent width of \mbox{C\,{\sc iv}} 5808 {\AA} versus \mbox{He\,{\sc ii}} 4686 {\AA}. From Table 1 we see that $\log\,EW(\mbox{C\,{\sc iv}}\,5808\,{\AA})$ = 1.38, while $\log\,EW(\mbox{He\,{\sc ii}}\,4686\,{\AA})$ = 1.83. 

Concentrating for the moment on the C emission lines, we see that the ratio of \mbox{C\,{\sc iv}} 5808 {\AA} to \mbox{C\,{\sc iii}} 5696 {\AA} = 4.3, consistent with a WC6 subclass \citep{cro98}. The ratio of \mbox{He\,{\sc ii}} 5411 {\AA} to \mbox{He\,{\sc i}} 5876 {\AA} is in the range of 8 to 10, (with the latter line possibly truncated by the NaI D line); this is consistent with a WN5 subclass \citep{smi96}. The strongest emission in the spectrum is of Figure 2 is \mbox{N\,{\sc iv}}, thus we classify this transition star as WN5/WC6, with an estimated error of plus or minus one subclass for each of the WN and WC subclassifications.

\begin{figure}
\centerline{\includegraphics[width=0.99\columnwidth]{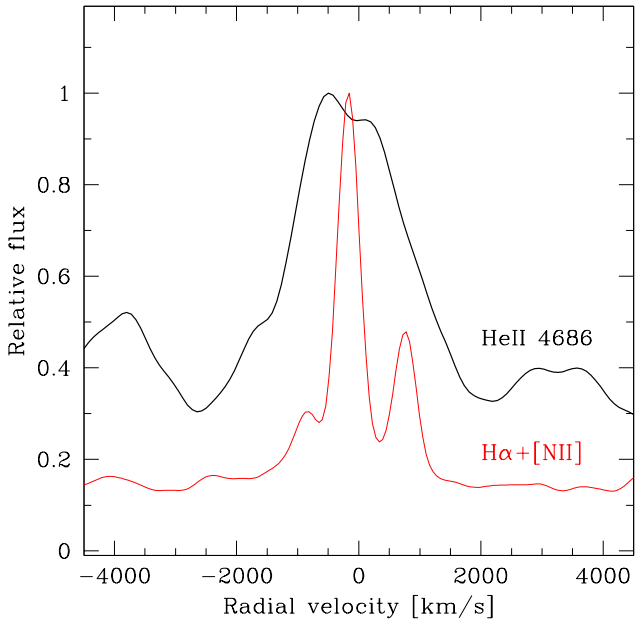}}
\caption{Comparison of the \mbox{He\,{\sc ii}\, 4686} and H$\alpha$ + \mbox{[N\,{\sc ii}]} emission line profiles.}\label{profile}
\end{figure}

The dereddened $V_0=20.0$ and $I_0=19.0$ combined with the true distance modulus $m-M=24.47$ \citep{vil06} result in the absolute magnitudes of  M31 84-1 $M_{\rm V}=-4.5$ and $M_{\rm I}=-5.5$, respectively. These values are consistent with typical magnitudes of WC6 and WN5 stars.

We note that there does remain the possibility that this star is a binary, composed of WC6 and WN5 components. Figure 1 shows that the angular separation of such a binary must be less than 0.1'', corresponding to almost 0.4 pc at the distance of M31. Time resolved spectroscopy might show the carbon and nitrogen lines changing their radial velocities in phase (cf WR 145 = MR 111 \citep{mas89}), proving that the new WR star is a transition WR object in a binary. Alternately, the carbon and nitrogen lines might move out of phase, proving that we have detected a binary WR star with two WR components. No radial velocity shifts of either element's lines would be indeterminate, consistent with a single transition star, or a very long period binary WR star.

\subsection{The new WN/C star's environment, location in M31 and metallicity}

In Fig.~\ref{profile} we show the profiles of the strong emission line \mbox{He\,{\sc ii}}\,4686\,{\AA} and the region around H$\alpha$, scaled vertically so that their maxima overlap. The helium line has a FWZI of 4000 km/s, typical of WR stars. This 4000 km/s FWZI is matched and confirmed by that of the \mbox{He\,{\sc ii}}\,5411\,{\AA} (see Table1), demonstrating that blending from other lines is not the cause of the large FWZI. The emission line of H$\alpha$ is much narrower than the helium 4686 and 5411 emission lines, as are the flanking lines of \mbox{[N\,{\sc ii}]} that are present. This, together with the other narrow Balmer lines, and the equally narrow \mbox{[S\,{\sc ii}]} and \mbox{[O\,{\sc iii}]} lines, which all show FWHM around 500 km/s (see Table 1), demonstrates that the WR star is immersed in a gas of low density and moderately high excitation. While it is possible that part of these narrow emission lines are diffuse interstellar gas, we note that in Fig.~\ref{FC}, (where the WN/C star labelled WR is clearly seen to be in emission), the star is unresolved in the narrowband F658N and F502N filters, relative to nearby stars, which do not show similar emission. We also see no trace of diffusive emissivity beyond the radius of the point-spread function of the WR star. We thus suggest that the new WN/C star is surrounded by its own emission-line nebula.

As noted above, theory predicts that transition stars will be very short-lived if their metallicity is unusually high. Should we be surprised to find such a star in M31, whose central regions' metallicity is higher than solar? M31 84-1 is at a deprojected M31 galactocentric distance of 52.1 arcmin (calculated using the same method as \citet{san12}), which corresponds to 11.8 kpc, assuming that the distance to M31 is 780 kpc; \citep{vil06}. It is inside the Population I ring located between 9 and 15 kpc from the center of M31, which contains the most active star formation regions in M31, and the vast majority of its WR stars (see figure 6 of \citet{neu12}).
The average radial velocity derived from the narrow emission lines, $\varv_{\rm rad}=-169 \pm 14\, \rm km\,s^{-1}$, is consistent with the rotational velocity of M31 \citep{che09} at the galactic position of M31 84-1.

We can estimate the metallicity in the environs of M31 84-1 by using the radial oxygen and nitrogen abundance profiles of M31 \citep{san12}. The average metallicity values at an M31 galactocentric distance of 11.8 kpc are: $\log(\rm O/H)+12 \sim 8.9$ is $\sim$ 0.2 dex higher, and $\log (\rm N/H+12) \sim 7.5$ is $\sim$ 0.3 dex lower, respectively, than the solar values of $8.69 \pm 0.05$, and $7.83 \pm 0.05$ \citep{asp09}. Furthermore, \citet{san12} demonstrated that there is significant intrinsic scatter around the observed M31 abundance gradient, with as much as $\sim 3$ times the systematic uncertainty in the strong-line diagnostics that they use. We conclude that the \citet{san12} determinations of metallicity at the galactocentric distance of M31 84-1 are consistent with solar metallicity. 

\subsection{WR stars in M31}

\citet{neu12} presented evidence that they have found at least 95\% of the unreddened WR stars in M31. Our survey is complementary to theirs because, as noted above, we focus on candidates with strong H$\alpha$ emission in images of the publicly-available LGGS survey \citep{mas06} which are quite red ($V - I \ga 2.0$). Thus we cannot detect new, unreddened WR stars in M31, but we are sensitive to reddened WR stars if immersed in H$\alpha$ nebulosity. We have detected only one new WR star out 441 red candidates with strong H$\alpha$ emission ($m(\rm H\alpha) - R \la -1.0$), and the total number of such objects in the LGGS images of M31 is at most a few thousand. This might suggest that at most $\sim10$ new, reddened WR stars remain to be found in M31. However, Fig.~\ref{FC} demonstrates how crowding makes it very difficult, with ground based imagery, to locate stars even with strong H$\alpha$. If M31 84-1's H$\alpha$ emission were just 2\% weaker we would have missed detecting it as a candidate. Most WR stars' H$\alpha$ emission is weaker than that of our new WR star. Furthermore, WR stars on the far side of the M31 disc may be so reddened as to be undetectable in LGGS imagery. We thus cannot reliably predict the number of highly reddened WR stars in M31, but the number may be significant.  

%==================================================================

\section{Conclusions}\label{conclusions}

We presented the observed and dereddened spectra of, and discussed the first likely transition WN/C transition WR star detected in M31. 
The coordinates of the new star are 00:43:43.61  +41:45:27.95 (J2000). It's spectral type is WN5/WC6, with an uncertainty of $\pm$1 in each spectral subtype. 
The star is located inside the Population I ring of star formation of M31, at a location which has metallicity comparable to that of the Sun. 
It is immersed in a hydrogen-rich, low density nebula of moderate excitation.
M31 84-1 demonstrates that a number of other, highly reddened WR stars probably remain to be found in M31.

%==================================================================
\section*{Acknowledgments}

This study has been supported in part by the Polish NCN grant
DEC-2013/10/M/ST9/00086.
We gratefully acknowledge the fine support at the MMT Observatory, and
The Local Group Galaxy Survey conducted at NOAO by Phil Massey and
collaborators.

This research has made use of the VizieR catalogue access tool, operated at CDS, Strasbourg, France.
Based on observations made with the NASA/ESA Hubble Space Telescope, obtained from the Data Archive at the Space Telescope Science Institute, 
which is operated by the Association of Universities for Research in Astronomy, Inc., under NASA contract NAS 5-26555.

Helpful suggestions from A. Moffat are gratefully acknowledged.

%==================================================================

\newpage

\label{lastpage}

\end{document}